\documentclass[12pt,epsf]{article}
\usepackage{graphicx, amsmath}

\begin{document}

\sloppy
\begin{flushright}{SIT-HEP/TM-42}
\end{flushright}
\vskip 1.5 truecm
\centerline{\large{\bf Modulated Inflation}}
\vskip .75 truecm
\centerline{\bf Tomohiro Matsuda\footnote{matsuda@sit.ac.jp}}
\vskip .4 truecm
\centerline {\it Laboratory of Physics, Saitama Institute of Technology,}
\centerline {\it Fusaiji, Okabe-machi, Saitama 369-0293, 
Japan}
\vskip 1. truecm
\makeatletter
\@addtoreset{equation}{section}
\def\theequation{\thesection.\arabic{equation}}
\makeatother
\vskip 1. truecm

\begin{abstract}
\hspace*{\parindent}
We have studied modulated inflation that generates curvature perturbation
from light-field fluctuation.
As discussed in previous works, even if the fluctuation of the inflaton
 itself does not generate the curvature perturbation at the horizon
 crossing, fluctuation of a light field may induce fluctuation for the
 end-line of inflation and this may lead to generation of cosmological
 perturbation ``at the end of the inflation''. 
Our scenario is different from those that are based on the fluctuations
of the boundary of the inflaton trajectory, as clearly explained in this
 paper by using the $\delta N$ formalism. 
In this paper, we will consider the perturbation of the inflaton
 velocity that can be induced by a light field other than the inflaton.
We also explain the crucial difference from the standard multi-field
 inflation model.
We show concrete examples of the modulated inflation scenario in which
non-gaussianity can be generated. 
We also discuss the running of the non-gaussianity parameter.
\end{abstract}

\newpage
\section{Introduction}
\hspace*{\parindent}
During inflation the vacuum fluctuations of all light scalar fields
${\cal M}_i$ 
are unstable and appear as classical random Gaussian inhomogeneities
with an almost scale-free spectrum of amplitude
$\delta {\cal M}_i\simeq H_I/2\pi$, where $H_I$ is the Hubble parameter
during inflation.
Inflation significantly stretches the wavelength of the fluctuations
over the Hubble horizon after inflation. 
Thus one can relate $\delta
{\cal M}_i$ 
in many different ways to the cosmological curvature perturbation
observed in the present Universe. 
The standard form for the number of
e-foldings elapsed during inflation is given by 
\begin{equation}
\label{original_1}
N =\frac{1}{M_p^2}\int^{\phi_N}_{\phi_e} \frac{V}{V_\phi}d\phi,
\end{equation}
where $\phi_N$ and $\phi_e$ are the values of inflaton field $\phi$,
corresponding to $N$ e-foldings and the end of inflation, respectively.
In fact, if there is fluctuation of a common spectrum 
$\delta \phi_N = H_I/2\pi$,
we can calculate the spectrum of the density perturbation
by using $\delta N$-formalism: 
\begin{equation}
 \delta_H^2 = \frac{4}{25}(\delta N)^2 
=\frac{4}{25}\left(\frac{V}{M_p^2V_\phi}\frac{H_I}{2\pi}\right)^2,
\end{equation}
which reproduces the standard result; the density perturbation generated
during single-field inflation.\footnote{In this paper we follow
the notations given in the textbook\cite{Lyth-book}.}
More generically, one may however expect several scalar fields playing
similar roles during inflation.
Assuming $\phi_e$ depends on such a light field, one encounters an
alternative mechanism for generating curvature perturbation.
This possibility has been discussed by Bernardeau et
al.\cite{modulated_hybrid} for modulated couplings in hybrid
inflation,\footnote{The word ``modulated fluctuations'' was introduced
by Kofman in Ref.\cite{Kof-1}}
and by Lyth\cite{At_the_end_of} for a multi-field model of
hybrid inflation, and more recently, by us\cite{at_the_end_of_trapped}
for trapping inflation combined with inhomogeneous preheating.
The multi-field model\cite{At_the_end_of} has been applied to brane
inflation in a throat as a solution to the serious $\eta$-problem in
string theory\cite{brane-deltaN}. 
In fact, finding a light inflaton field whose mass is protected by
symmetry is rather hard in string theory models, especially when masses
are determined by calculable mechanism of moduli stabilization.
It is therefore very important to find a light field other than the
inflaton, which can contribute to the curvature perturbation. 
Moreover, it is obvious that the standard single-field inflationary
scenario cannot explain large ($f_{nl} \ge O(1)$)
non-gaussianity\cite{large-non}. 
The clear difference from the standard multi-field inflation is
explained in Appendix B.
There are many ``alternatives'' in which such light fields
play crucial roles in generating the cosmological perturbation, 
such as cosmological perturbation generated (1) long after
inflation (curvatons)\cite{curvaton_1,curvaton_low}, (2) during
preheating (inhomogeneous preheating)\cite{inhomogeneous_preteating}, or
(3) during reheating (inhomogeneous
reheating)\cite{inhomogeneous_reteating}. 
Moreover, by combining these ideas it is possible to generate the initial
perturbation of the curvaton from inhomogeneous
preheating\cite{curvaton_from_PR}.\footnote{Large non-gaussianity can be
generated during inhomogeneous preheating.
Even if significant non-gaussianity is not generated at preheating,
small ratio $r\equiv \rho_{\chi}/\rho_{total} \ll 1$ at the decay can
lead to large non-gaussianity, as has been discussed in
Ref.\cite{curvaton-fnl} for curvatons. Here $\rho_{\chi}$ is the energy
density of the preheat field.
Note that $f_{nl}$ in these inhomogeneous preheating scenarios can take
either sign.} 
% added-footnote-kokomade
 
In summary, the standard inflationary scenario uses genuine inflaton
fluctuation $\delta \phi_N$ to generate the curvature perturbation
during inflation, while in recent works\cite{modulated_hybrid,
At_the_end_of} where perturbation is generated ``at the end of
inflation'',  curvature perturbation is generated from $\delta\phi_e$ 
that is indirectly generated by fluctuation of another light
field.\footnote{Note that in other scenarios\cite{curvaton_1,
curvaton_low, inhomogeneous_preteating, inhomogeneous_reteating,
curvaton_from_PR} dynamics ``after'' inflation is used to generate
curvature perturbation. In this paper, we however focus our attention on
scenarios that can work ``during'' inflation.}
In both cases, the fluctuation of the boundary condition leads to
$\delta N$.
Looking at the original equation (\ref{original_1}), one may however
find that fluctuations induced by other components (i.e., $M_p^{-2}$ or
$V/V_\phi$) may generate curvature perturbation if these components are
modulated during inflation due to their dependence on a light
field(moduli).  
%% added after ver.3 begin
More precisely, we will consider the perturbation of the
velocity of the inflaton field, which is generated by the fluctuation of
the moduli.
The number of e-foldings for multi-field inflation is given by
\begin{equation}
N = \int H\frac{\dot{\phi}d\phi +\dot{\cal M}
d{\cal M}}{\dot{\phi}^2 + \dot{\cal M}^2}.
\end{equation}
If there is no bend in the trajectory and $\dot{\cal M}\simeq 0$,
the perturbation related to the inflaton velocity can be expanded as 
\begin{equation}
\label{delta-vel}
\delta N 
\simeq -\int^{\phi_N}_{\phi_e}\frac{H}{\dot{\phi}^2}\left(
\delta\dot{\phi}-\dot{\phi}A\right)d\phi,
\end{equation}
where we consider linear scalar perturbations of the metric,
\begin{equation}
ds^2 = -(1+2A)dt^2 + 2aB_idx^idt + a^2\left[(1-2\psi)\delta_{ij}+2E_{ij}
\right]dx^i dx^j.
\end{equation}
If the kinetic term is standard,\footnote{If the inflaton kinetic term
depends on the moduli field, there is a contribution that remains constant
after horizon crossing\cite{modulated-kinetic}.} the perturbation related to
the 
inflaton velocity (\ref{delta-vel}) is proportional to $k^2/a^2$ and
decreases after horizon crossing.
Therefore, we need to introduce a factor $e^{-2Ht}$ in the calculation,
if the perturbation is not due to a moduli-dependent kinetic term.
Because of the integration, the contribution may not vanish at a large
distance.
The simplest example would be a perturbation
$\delta c \times e^{-2Ht}$, which leads to $\int  \delta c H
e^{-2Ht}dt\simeq -\frac{\delta c}{2}$, while for non-decaying
perturbation $\delta c$, the integral leads to $\int H \delta c dt
\simeq N\delta c$.
%% added after ver.3 end
Based on this simple idea, in this paper 
we consider an alternative
mechanism for generating the curvature perturbation that relies
neither on $\delta \phi_N$ nor $\delta \phi_e$.
This is the crucial difference from the previous scenarios\cite{modulated_hybrid}.

In Sec.2, we first consider hybrid inflation with a moduli-dependent
inflaton mass $m^2({\cal M})\equiv m^2_0(1+\beta\log({\cal M}/M_*))$,
which induces fluctuation related to $(V/V_\phi)$.
Then in Sec.3, we examine the possibility of generating the curvature
perturbation in the brane inflation model with a modulated
coupling. 
Note that our examples are based on typical moduli-dependences
in conventional models, although we will not specify the model because
of the generality of our argument.
In Sec.4, we consider the generation of the cosmological perturbation
from modulated fluctuation of the effective Planck scale. 
The possibility discussed in Sec.4 is in a sense very natural, because
the Planck scale always appears in the equation. 
In some specific examples
light fields can be identified with volume of extra dimensions that
evolves slowly during inflation. 
In an appendix, we consider MSSM
inflation that has been advocated by Allahverdi
et. al.\cite{MSSM-inflation}. 
In this appendix, we consider a rather peculiar source for the
moduli-dependence, the fluctuation of $\phi_0$ that denotes the point
where the secondary minimum appears. 
Our argument for MSSM inflation is that if 
couplings depend on moduli, the modulation can be mediated to $\phi_0$ 
that determines the number of e-foldings. 

We calculate explicit forms of the non-gaussianity
parameter $f_{nl}$ and show how non-gaussianity can be generated.
Conclusions and discussions are presented in Sec.5.

\section{Modulated hybrid inflation}
\hspace*{\parindent}
Let us start with the original hybrid inflation model. 
Hybrid inflation has the effective potential for the two fields
$(\phi,\sigma)$,
\begin{equation}
V(\phi,\sigma)=\lambda\left(\sigma^2-v^2\right)^2
+\frac{1}{2}g^2\phi^2\sigma^2 + V(\phi),
\end{equation}
where $\phi$ is the inflaton and $\sigma$ is the trigger field.
In this section we consider a specific example of the model, whose
inflaton potential is given by $V(\phi)=m^2\phi^2/2$.
Here the end of inflation expansion occurs at
\begin{equation}
\phi_e=\frac{\sqrt{\lambda}v}{g}.
\end{equation}
The number of e-foldings is given by
\begin{equation}
N=\frac{\lambda v^4}{M_p^2 m^2}\log\frac{\phi_N}{\phi_e}
=\frac{1}{\eta_\phi}\log\frac{\phi_N}{\phi_e},
\end{equation}
where the definition of the slow-roll parameter is $\eta_\phi\equiv
m^2/3H_I^2=m^2 M_p^2/(\lambda v^4)$.
Modulated fluctuations of couplings $\lambda({\cal M})$ or $g(\cal M)$
are discussed in Ref.\cite{modulated_hybrid}, which may (or may not)
lead to the fluctuation $\delta \phi_e$, and therefore to 
indirect generation of $\delta N$ at the end of inflation.

On the other hand, since modulated fluctuation of $m^2$ does not lead to
$\delta \phi_e$, $\delta N $ is not generated from modulated $m$
at least at the end of inflation.
Alternatively, it leads to another kind of the fluctuation 
\begin{equation}
\delta N_{\cal M} \propto
\left(\frac {m'}{m}\right)\delta {\cal M},
\end{equation}
where $m'$ is the derivative of $m$ with respect to ${\cal M}$.

We consider a concrete example of the moduli-dependent mass
\begin{equation}
m^2({\cal M})\equiv m^2_0\left[1+\beta\log({\cal M}/M_*)\right],
\end{equation}
where $\beta$ and $M_*$ are model-dependent parameters.\footnote{Here 
for simplicity we neglect loop corrections related to $\phi$.
Of course this does not destroy the above argument as far as the
original hybrid inflation scenario is successful.}
This leads to the fluctuation of the inflaton velocity.
Considering the exponential factor, we find
\begin{equation}
\delta N_{\cal M} \simeq \beta \left(\frac {\delta{\cal M}}{{\cal M}}\right).
\end{equation}
The inflaton fluctuation may be negligible if inflation is
fast-roll\cite{fast-roll-inflation} and where inflaton mass is as large
as $m\sim H$.
In this specific case, the perturbation generated by the modulated mass
can dominate the curvature perturbation.
However, in other cases, inflaton fluctuation would be significant.
The modulated perturbation can thus dominate when the condition
\begin{equation}
\left(\frac{\cal M}{\phi_N}\right)< \beta\eta_\phi
\end{equation}
is satisfied.\footnote{See appendix C for the case with $p\equiv \dot{\cal
M}/\dot{\phi} \gg 1$.}
Moreover, the mass of ${\cal M}$ (i.e., $m_{\cal M}$) must be less than
$H_I$ during inflation, which leads to an 
another condition given by
\begin{equation}
\label{another-slow}
m^2_{\cal M}\simeq \beta m^2_0\left(\frac{\phi_N}{\cal M}\right)^2
< H_I^2.
\end{equation}
Combining these equations and assuming that the amplitudes for
fluctuations are comparable ($\delta {\cal M} = \delta \phi_N$), we
obtain\footnote{See also Ref.\cite{hilltop-curvaton, dine-rand-thom}
for natural values of $\eta$-parameter.}
\begin{equation}
\label{final-1}
\beta \eta_\phi > 1.
\end{equation}

It is very important to calculate a non-gaussianity parameter of our
modulated 
inflation scenario, since the non-gaussianity parameter is expected to
distinguish the curvature perturbation generated by alternative
mechanisms from the one that is generated by conventional inflaton
fluctuation. 
Here we consider the definition of non-gaussianity parameter $f_{nl}$
through curvature perturbation\cite{Lyth-book}.
By using the $\delta N$-formalism, we can write
\begin{equation}
\zeta = N'\delta {\cal M} +\frac{1}{2}N'' (\delta{\cal M})^2,
\end{equation}
where higher terms are dropped because they are not important here.
The explicit form of the non-gaussianity parameter is then given by
\begin{equation}
\label{f_nl_form}
-\frac{3}{5}f_{nl}\equiv \frac{1}{2}\frac{N''}{(N')^2},
\end{equation}
where the prime denotes the derivative with respect to the moduli.
Applying the explicit form of $m^2({\cal M})$ to Eq.(\ref{f_nl_form}),
we find for $|N\beta|\ll 1$,
\begin{equation}
f_{nl}=-\frac{5}{6}\frac{N''}{(N')^2}\sim
\frac{1}{ \beta},
\end{equation}
which can be $f_{nl}\sim O(1)$ if the inflaton is fast-rolling.
Here $f_{nl}$ may take either sign.
A running of non-gaussianity may hopefully be
used to distinguish modulated inflation from other models of large
non-gaussianity. 
The explicit form of the running of the non-gaussianity at a large
distance is\cite{Lyth-book}
\begin{equation}
\frac{d \ln f_{nl}}{d \ln k}\simeq 0.
\end{equation}

\section{Modulated brane inflation}
\hspace*{\parindent}
Brane inflation is one of the specific accomplishments of the inflationary
universe within the brane world framework in the string theory.
Inflation potential during brane inflation is typically given by
\begin{equation}
V(\phi)=\frac{1}{2}\gamma H_I^2 \phi^2 + V_0\left(1-\lambda
\frac{M^n}{\phi^n}\right),
\end{equation}
where $\gamma$, $\lambda$ and $M$ are model-dependent parameters.
For simplicity we consider the case in which only $\lambda$ depends on the
light moduli.
Here $|\gamma|\ll 1$ is required to make enough e-foldings during
inflation.
The inflaton field $\phi$ measures the brane distance, and 
the end of inflation is induced by brane collision that occurs when 
two branes come closer than the string scale.
Note that the situation in this inflationary model is identical to
 the conventional hybrid inflation model in so far as the first term
(mass term) dominates the evolution of the inflaton field.
Therefore, modulated inflation may occur in the same way
as the conventional hybrid inflation
if the mass term dominates the inflaton evolution.
On the other hand, if the second term (Coulomb-like potential) dominates
the inflaton potential, modulation will appear through the dimensionless
coupling constant $\lambda$ and the situation of modulated inflation
will be different.
Again, inflaton fluctuation leads to 
\begin{equation}
\delta N_\phi = N(n+2)\times \frac{\delta \phi_N}{\phi_N},
\end{equation}
while modulation related to moduli-dependent coupling $\lambda({\cal
M})$ leads to
\begin{equation}
\delta N_{\cal M} \simeq \frac{\lambda'}{2\lambda}\delta {\cal M},
\end{equation}
where $\lambda'$ is the derivative with respect to ${\cal M}$.
For a concrete example let us consider 
$\lambda({\cal M}) \equiv \lambda_0 e^{-c {\cal M}/M_*}$ during
inflation, where 
$c$ is a dimensionless constant and $M_*$ is a scale parameter.
Values of these parameters are model dependent.
Then we obtain
\begin{equation}
\delta N_{\cal M} \simeq c \times \frac{\delta {\cal M}}{2M_*}.
\end{equation}
To take into account the condition where the modulated perturbation
exceeds the conventional cosmological perturbation originating from
inflaton fluctuation, we 
consider  the condition 
$\delta N_\phi \ll \delta N_{\cal M}$ that leads to
\begin{equation}
\label{condition-brane}
\frac{\phi_N}{M_*}\gg N\frac{n+2}{c}.
\end{equation}
Another condition is needed to ensure that moduli field ${\cal M}$ is
light during inflation.
The effective mass induced by the inflaton potential is
\begin{equation}
m_{\cal M}^2 \simeq 
\lambda V_0 \frac{M^n}{\phi_N^n}\left(\frac{c}{M_*}\right)^2,
\end{equation}
which must be less than the Hubble parameters during inflation,
$H_I \equiv \frac{V_0}{3M_p^2}$.
We conclude that our modulated inflation scenario works with
typical brane inflation potential provided that there is a light field
which satisfies the above conditions.

Again, a non-gaussianity parameter is important in distinguishing
curvature perturbations.
Applying a more generic form $\lambda({\cal M})\equiv \lambda_0
e^{\alpha({\cal M})}$ to Eq.(\ref{f_nl_form}), and assuming that the
dominant component during inflation is $\alpha({\cal M})
\simeq \alpha_n {\cal M}^n/M_*^n$, 
we find
\begin{equation}
f_{nl}=-\frac{5}{6}\frac{N''}{(N')^2}
\simeq\frac{1}{\alpha_n}\frac{M_*^n}{{\cal M}^n},
\end{equation}
which can be large and may take either sign.

\section{Inflation with modulated Planck scale}
\hspace*{\parindent}
The radial mode ${\cal M}_e$ of extra compact space may satisfy the
slow-roll condition if the effective potential during inflation
satisfies the slow-roll condition.
In this case, there can be $\delta N$ perturbation generated by the
modulated Planck scale during inflation,
\begin{equation}
\delta N = -2 N \times \left(\frac{M_p'}{M_p}\right)\delta {\cal M}_e,
\end{equation}
where $M_p'$ denotes the derivative of $M_p$ with respect 
to ${\cal M}_e$.\footnote{A non-trivial kinetic term may cause deviation
from the standard gaussian perturbation, which leads to
another kind of non-gaussianity\cite{kin-fnl}.
However, the result will be highly model-dependent. For simplicity,
the kinetic term is assumed to be minimal.} 
A specific example of the slow-roll condition for the
radion is discussed in Ref.\cite{cline-slow-extra}.
The condition given in Ref.\cite{cline-slow-extra} is apparent in the
original frame, but in the Einstein frame the rescaling of the
inflaton potential is crucial.
%added
For example, in Brans-Dicke theory,
the exponential factor induced by the metric rescaling can be canceled by
the scalar-field rescaling if the potential is quartic.
This cancellation may happen if the inflaton potential is given by 
$V(\phi)=\lambda \phi^4/4 $ for chaotic inflation.
%added
Note that the rescaling of the inflation potential is highly
model-dependent and also may be shifted by the quantum effect.
In this respect, one cannot simply assume that the ``constant'' vacuum
energy during inflation is not affected by the rescaling,
especially when the scale is generated dynamically.
In any case, it is however possible to assume slow-roll evolution of the
radial mode ${\cal M}_e$ during inflation that rapidly settles to
the minimum after inflation\cite{cline-slow-extra}.

Note that our mechanism for modulated Planck scale works in that
the Planck scale determines the number of e-foldings.
The required condition for generating the curvature perturbation
is the slow-roll condition for the corresponding moduli field, which is 
 highly model-dependent.

\section{Conclusions and discussions}
\hspace*{\parindent}
We have studied a new class of modulated fluctuations that generates
curvature perturbation from light-field fluctuation.
According to Ref.\cite{modulated_hybrid} and \cite{At_the_end_of},
fluctuation of a light field may induce fluctuation of the end-point of
inflation expansion ($\delta \phi_e \ne 0$), which may lead to
generation of cosmological perturbation at the end of the inflation.
Contrary to these previous scenarios of modulated fluctuations, the origin
of the cosmological perturbation in our scenario is not $\delta \phi_e$.
As far as we know, this is the first realization of a
modulated scenario without $\delta \phi_e$.
Moreover, we showed a concrete example of modulated inflaton in which 
large non-gaussianity can be generated.
The possibility of large non-gaussianity has been suggested in
Ref.\cite{modulated_hybrid}, but there was no specific
example.\footnote{Large non-gaussianity in the multi-field model 
of hybrid inflation has been discussed in Ref.\cite{At_the_end_of} as
a specific example.}
As we have discussed in Ref.\cite{hilltop-curvaton} for curvatons, it is
rather easy to generate the required value of spectral index in these
alternative scenarios, since in these models the form of the light-field
potential is not constrained by the kinematic requirements of the 
inflation expansion. 
In fact, in hilltop scenarios\cite{hilltop-inflation} of these 
alternatives\cite{hilltop-curvaton}, the spectral index may be related
to the negative $\eta$-parameter of a light field during inflation,
which can be justified in conventional supergravity 
models\cite{hilltop-curvaton, dine-rand-thom, hilltop-inflation}.

An obvious deficit of modulated scenarios is the famous moduli problem,
as these scenarios always require moduli fields displaced from their
true minimum at least at the time when the perturbation is generated.
Late-time entropy production such as thermal inflation\cite{Lyth-book}
 may solve this
problem, but thermal inflation may not work if the energy scale of the 
primordial inflation is very low.
Of course it is extremely hard to build an inflationary model at low
scales\cite{low-infla}\footnote{See also Ref.\cite{curvaton_low} for the
condition for the inflation energy scale in the curvaton scenario.}, 
but low-scale inflation may become very
important if the gravitational effect is observed in the Large Hadron
Collider(LHC). 
Moreover, many inflationary models that are based on single
$\delta \phi_N$-fluctuation may be excluded if large non-gaussianity is
confirmed by the observation\cite{large-non}. 
Note that the moduli problem does not appear in multi-field models such
as Ref.\cite{At_the_end_of, brane-deltaN}
or inhomogeneous preheating scenarios\cite{inhomogeneous_preteating},
in which a non-gaussianity parameter can be large and at the same time,
low-scale inflation is possible.\footnote{See also
Ref.\cite{fnl_from_PR_other}.}

\section{Acknowledgment}
We wish to thank K.Shima for encouragement, and our colleagues at
Tokyo University for their kind hospitality.
\appendix

\section{Modulated MSSM inflation}
In this appendix, we show another possibility of introducing
moduli dependence to the cosmological perturbation theory.
We start with the Minimal Supersymmetric Standard Model (MSSM), which 
 is a well motivated extension of the Standard Model(SM).\footnote{MSSM
 inflation\cite{MSSM-inflation} is an attractive idea, but it may be
 excluded if large non-gaussianity is confirmed by the
 observation\cite{large-non}. One way to solve this problem is to
 generate cosmological perturbation by using modulated inflation.} 
Considering a flat direction $\phi$ with non-renormalizable
superpotential 
\begin{equation}
W=\frac{\lambda_n}{n}\frac{\Phi^n}{M_p^{n-3}},
\end{equation}
where $\Phi$ denotes the superfield related to $\phi$.
We find the scalar potential
\begin{equation}
 V=\frac{1}{2}m^2\phi^2+A\cos (n\theta+\theta_A)
\frac{\lambda_n \phi^n}{nM_p^{n-3}}+
\frac{\lambda_n^2\phi^{2(n-1)}}{M_p^{2(n-3)}},
\end{equation}
where $m$ and $\theta_A$ come from the soft supersymmetry(SUSY) breaking
mass and the A-term, respectively.
Here $\phi$ and $\theta$ denote the radial and angular coordinates of
the scalar component of the superfield $\Phi$.
The potential has a secondary minimum at 
\begin{equation}
\phi_0\sim \left(\frac{m M_p^{n-3}}{\lambda_n}\right)^{1/(n-2)},
\end{equation}
but the potential barrier, however, disappears if the coefficient of the
A-term ($A$) satisfies the condition
\begin{equation}
A^2=8(n-1)m^2.
\end{equation}
Around this secondary minimum with the coefficient 
$A^2\simeq 8(n-1)m^2$, the field only feels the third derivative of the
potential.
Then inflation may start near $\phi_0$, which leads to the number of
e-folds
\begin{equation}
N \simeq \frac{\phi_0^3}{2n(n-1)M_p^2(\phi_0-\phi)}.
\end{equation}
As we are interested in fluctuation related to $\phi_0({\cal M})$, we
calculate the derivative of $N$ with respect to ${\cal M}$;
\begin{equation}
N_{\cal M}\simeq 3 \frac{\phi_0'}{\phi_0}-\frac{\phi_0'}{\phi_0-\phi}
\sim -\frac{\phi_0'}{\phi_0-\phi}.
\end{equation}
Here $|\phi_0-\phi|\ll |\phi_0|$ is considered.
See Ref.\cite{MSSM-inflation} for more details.
The amplitude of the perturbation is therefore given
by\footnote{Numerical factors are neglected for simplicity.}
\begin{equation}
\delta N_{\cal M} \simeq 
\label{modulated-mssm}
 N \frac{m M_p}{\phi_0^2}\phi_0',
\end{equation}
where the amplitude is assumed to be 
\begin{equation}
\delta {\cal M}\simeq H_I/2\pi \simeq \frac{V(\phi_0)^{1/2}}{2\pi M_p}
\simeq \frac{(n-2)}{2\pi\sqrt{2n(n-1)}M_p}m\phi_0.
\end{equation}
Comparing our result (\ref{modulated-mssm}) with the standard result,
which has been obtained from the fluctuation of the inflaton
field\cite{MSSM-inflation}, we obtain the required
condition  
\begin{equation}
\label{MSSM-condition}
\phi_0' \propto 
\phi_0 \frac{\lambda_n'}{\lambda_n}\gg 1,
\end{equation}
which is needed for $\delta N_{{\cal M}}$ to dominate the cosmological
perturbation. 
The slow-roll condition for the light field ${\cal M}$ may put a
severe restriction (lower bound) on the inflation energy scale.
Note that MSSM inflation requires cancellation of the second derivative
of the inflaton field, which makes it possible to construct a low-scale
inflation model.
On the other hand, the naive estimation of the moduli
mass is $O(m)$, if the mass is not protected.
In fact, introducing explicit moduli-dependence to MSSM action is
not easy at this moment, especially when we need to introduce
 a moduli field
whose mass can be much smaller than the SUSY-breaking
mass during inflation.\footnote{$\lambda_n$ may be generated from
instantons in intersecting brane model. In this
case, the moduli may be related to the area bounded by branes, which 
 may be light during inflation\cite{intersect-brane}.}

\section{Source term for the curvature perturbation}
The simplest way to see the source term for the curvature perturbation in
modulated inflation is to consider the evolution of the curvature
perturbation  
\begin{equation}
\dot{\cal R}=-H\frac{\delta P}{\rho+P},
\end{equation}
where $\delta P$ is the pressure perturbation.
The key idea in the modulated inflation scenario is the 
explicit ${\cal M}$-dependence in the ``slow-roll velocity'';
$\dot{\phi}=V_\phi/3H_I$.
Here we consider hybrid inflation that leads to 
$\dot{\phi}=m({\cal M})^2 \phi/3H_I$.
The modulated perturbation thus leads to 
\begin{equation}
\delta \dot{\phi} \simeq 2\dot{\phi}\frac{m'}{m}\delta {\cal M},
\end{equation}
which leads to the pressure perturbation
\begin{equation}
\delta P \simeq \dot{\phi}\delta \dot{\phi} \simeq 
2\dot{\phi}^2\frac{m'}{m}\delta {\cal M}.
\end{equation}
From the integral of $\dot{\cal R}$ and the exponential factor, we find
the contribution from the modulated inflation;
\begin{equation}
\label{curv-mud}
\Delta {\cal R} \simeq  \frac{m'}{m}\delta {\cal M},
\end{equation}
where $\dot{\cal M}\simeq 0$ is assumed during inflation.

On the other hand, the term proportional to $k^2/a^2$ has been
disregarded  
in the standard equation for the multi-field inflation\cite{two-field}.
Obviously, ``bent'' in the inflation trajectory 
is not important in modulated inflation, but the correction from the
decaying term can be significant after integration.
For example, considering flat potential that leads to $\dot{\cal M}=0$,
we find the curvature perturbation (\ref{curv-mud}),
while there is no ``bent'' in the inflation
trajectory.
The key is the explicit ${\cal M}$-dependence of
the inflaton velocity $\dot{\phi}({\cal M})$, which has been disregarded
in the standard calculation of multi-field inflation.
Note that the source term that is proportional to $\dot{\theta}$ does
not generate 
a significant correction to the curvature perturbation if there is no
significant bend in the trajectory.

\section{Modulated inflaton for $p\equiv \dot{\cal M}/\dot{\phi}\gg 1$}
In appendix B, we considered only the limiting case for $p\ll 1$.
Note that $\dot{\cal M}\simeq 0$ represents the ideal situation for the
modulated inflation and is very useful to explain the origin of the
curvature perturbation in modulated inflation.
However, to attain
$\dot{\cal M} = 0$ we have to consider a fine-tuning(i.e. careful
cancellation
with an additional potential for ${\cal M}$). 
Moreover, considering a more general situation the
energy landscape during inflation can 
be changed by the motion of ${\cal M}$, and therefore can participate in
the inflaton trajectory.  
In this respect, the significance of modulated inflation should be
considered for arbitrary $p$.
In this appendix, we discuss what happens for $p\gg 1$.

Let us consider the slow-roll velocity of the fields at horizon crossing:
\begin{eqnarray}
\dot{\cal M}\simeq \frac{V_{\cal M}}{3H_I}&\simeq&
 \frac{\beta m_0^2 \phi^2}{6H_I {\cal M}}\\
\dot{\phi}\simeq \frac{V_\phi}{3H_I}&\simeq&
 \frac{m^2 \phi}{3H_I},
\end{eqnarray}
which leads to the ratio
\begin{equation}
p \equiv \frac{\dot{\cal M}}{\dot{\phi}}\simeq \beta\frac{\phi}{2\cal M}.
\end{equation}
The curvature perturbation generated at the horizon crossing 
is given by\cite{two-field}
\begin{equation}
{\cal R}=H_I \left(\frac{\dot{\phi}Q_\phi + \dot{\cal M}Q_{\cal M}}
{\dot{\phi}^2 + \dot{\cal M}^2}\right) \simeq \frac{1+p}{1+p^2}
{\cal R}_0,
\end{equation}
where $Q_\phi$ and $Q_{\cal M}$ are the Sasaki-Mukhanov variables and
${\cal R}_0$ denotes the curvature perturbation for the conventional
hybrid-type inflation(i.e. for $p=0$).
For $p\ll 1$, we may disregard the effect of the field ${\cal M}$ in the
curvature perturbation generated at the horizon crossing. 
However, for $p \ge1$, we should consider the curvature
perturbation caused by the field ${\cal M}$ {\bf at the horizon
crossing.}

Let us consider modulated inflation for $p \gg 1$.
The perturbation related to the modulated inflation (i.e. the
perturbation caused by $\delta \dot{\phi}$ and $\delta \dot{\cal M}$)
 gives
\begin{equation}
\dot{\cal R}=-H\frac{\delta P}{\rho+P}
\simeq \frac{ -H}
{(1+p^2)}\left[\beta\frac{\delta {\cal M}}{\cal M} 
+2p^2  \frac{\delta {\phi}}{\phi}\right]e^{-2Ht}.
\end{equation}
Note that the second term related to $\delta \dot{\cal M}$ becomes
important for $p\gg 1$.  

Using the above results, we can examine the conditions for modulated
inflation.
For $p \gg 1$, the main source of the curvature perturbation at the
horizon crossing is caused by the motion of ${\cal M}$, which is 
multiplied by a factor $(1+p)/(1+p^2)\sim p^{-1}$.
A factor $1/(1+p^2)\sim p^{-2}$ appears for 
 modulated inflation caused by $\delta \dot{\phi}$, while the
factor is $p^2/(1+p^2)\sim 1$ for $\delta \dot{\cal M}$.
In addition to the above conditions, we must consider the condition
(\ref{another-slow}).
Finally, we conclude that modulated inflation can be significant
for $p\gg 1$, but in this case the roles played by $\phi$ and ${\cal M}$
are exchanged.


\begin{thebibliography}{1}
\bibitem{Lyth-book}
A.R. Liddle and D.H. Lyth,
{\it Cosmological inflation and large scale structure,
Cambridge, UK: Univ. Pr. (2000)}. 
\bibitem{modulated_hybrid}
F. Bernardeau, L. Kofman and J.P. Uzan, 
{\it Modulated fluctuations from hybrid inflation,
	Phys.Rev.D70(2004)083004} [astro-ph/0403315];
  F.~Bernardeau and J.~P.~Uzan,
  ``Inflationary models inducing non-gaussian metric fluctuations,''
  Phys.\ Rev.\  D {\bf 67}, 121301 (2003)
  [arXiv:astro-ph/0209330].
\bibitem{Kof-1}
L. Kofman
{\it Probing string theory with modulated cosmological fluctuations}
[astro-ph/0303614].
\bibitem{At_the_end_of}
D. H. Lyth,
{\it Generating the curvature perturbation at the end of inflation,
JCAP 0511:006,2005} [astro-ph/0510443].
\bibitem{at_the_end_of_trapped}
T. Matsuda,
{\it Cosmological perturbations from inhomogeneous preheating and
	multi-field trapping, JHEP 0707:035,2007} [arXiv:0707.0543].
\bibitem{brane-deltaN}
T. Matsuda,
{\it Elliptic Inflation: Generating the curvature perturbation without
slow-roll, JCAP 0609:003,2006} [hep-ph/0606137];
D. H. Lyth, Antonio Riotto,
{\it Generating the Curvature Perturbation at the End of Inflation in
	String Theory, Phys.Rev.Lett.97(2006)121301} [astro-ph/0607326].
\bibitem{large-non}
A.P.S. Yadav and B. D. Wandelt, 
{\it Detection of primordial non-Gaussianity in the WMAP 3-year
	data at above 99.5 percent confidence} [arXiv:0712.1148].
\bibitem{curvaton_1}
A. D. Linde and V. F. Mukhanov,
{\it Nongaussian isocurvature perturbations from inflation,
Phys.Rev.D56(197)535} [astro-ph/9610219];
D. H. Lyth, D. Wands,
{\it Generating the curvature perturbation without an inflaton,
	Phys.Lett.B524(2002)5} [hep-ph/0110002];
T. Moroi, T. Takahashi,
{\it Effects of Cosmological Moduli Fields on Cosmic Microwave
	Background, Phys.Lett.B522(2001)215} [hep-ph/0110096].
\bibitem{curvaton_low}
T. Matsuda,
{\it Curvaton paradigm can accommodate multiple low inflation scales,
Class.Quant.Grav.21(2004)L11} [hep-ph/0312058];
K. Dimopoulos, D. H. Lyth and Y. Rodriguez,
{\it Low scale inflation and the curvaton mechanism,
JHEP 0502:055,2005} [hep-ph/0411119];
Y. Rodriguez,
{\it Low scale inflation and the immediate heavy curvaton decay,
Mod.Phys.Lett.A20:2057,2005} [hep-ph/0411120].
\bibitem{inhomogeneous_preteating}
E. W. Kolb, A. Riotto and A. Vallinotto,
{\it Curvature Perturbations from Broken Symmetries,
Phys.Rev.D71(2005)043513}[astro-ph/0410546];
T. Matsuda,
{\it Generating the curvature perturbation with instant preheating,
JCAP 0703:003,2007} [hep-th/0610232];
T. Matsuda,
{\it Brane inflation without slow-roll, JHEP 0703:096,2007}
[astro-ph/0610402];
T. Matsuda,
{\it Generating curvature perturbations with MSSM flat directions,
JCAP 0706:029,2007}[hep-ph/0701024].
\bibitem{inhomogeneous_reteating}
G. Dvali, A. Gruzinov and M. Zaldarriaga, 
{\it A new mechanism for generating density perturbations from
	inflation, Phys.Rev.D69(2004)023505} [astro-ph/0303591];
F. Vernizzi,
{\it Cosmological perturbations from varying masses and couplings,
Phys.Rev.D69(2004)083526} [astro-ph/0311167];
K. Enqvist, A. Mazumdar and M. Postma,
{\it Challenges in generating density perturbations from a fluctuating
inflaton, Phys.Rev.D67(2003)121303} [astro-ph/0304187];
A. Mazumdar,
{\it A model for fluctuating inflaton coupling: (s)neutrino induced 
adiabatic, Phys.Rev.Lett.92(2004)241301} [hep-ph/0306026].
\bibitem{curvaton_from_PR}
T. Matsuda,
{\it Topological curvatons, Phys.Rev.D72(2005)123508}
[hep-ph/0509063];
T. Matsuda,
{\it Hybrid curvatons from broken symmetry, JHEP 0709:027,2007}
	[arXiv:0708.4098];
T. Matsuda,
{\it NO Curvatons or Hybrid Quintessential Inflation, JCAP 0708:003,2007}
	[arXiv:0707.1948].
\bibitem{modulated-kinetic}
T. Matsuda,
{\it Modulated inflation from kinetic term, JCAP 0805:022,2008} 
[arXiv:0804.3268].
\bibitem{curvaton-fnl}
D. H. Lyth, C. Ungarelli and D. Wands, 
{\it The Primordial density perturbation in the curvaton scenario,
Phys.Rev.D67(2003)023503}[astro-ph/0208055];
Q. G. Huang, 
{\it Large Non-Gaussianity Implication for Curvaton Scenario}
[arXiv:0801.0467].
\bibitem{MSSM-inflation}
R. Allahverdi, K. Enqvist, J. Garcia-Bellido, A. Mazumdar,
{\it Gauge invariant MSSM inflaton, Phys.Rev.Lett.97(2006)191304} 
[hep-ph/0605035];
A. Mazumdar,
{\it New developments on embedding inflation in gauge theory and
	particle physics} [arXiv:0707.3350];
R. Allahverdi, A. Kusenko and A. Mazumdar,
{\it A-term inflation and the smallness of the neutrino masses,
JCAP {0707}, 018 (2007)} [hep-ph/0608138];
R. Allahverdi, K. Enqvist, J. Garcia-Bellido, A. Jokinen and A. Mazumdar,
{\it MSSM flat direction inflation: slow roll, stability, fine tunning
	and reheating, JCAP {0706}, 019 (2007)} 
[hep-ph/0610134].
\bibitem{fast-roll-inflation}
A. Linde,
{\it Fast roll inflation, JHEP 0111:052,2001} [hep-th/0110195];
K. Dimopoulos, M. Axenides, 
{\it Hybrid inflation without flat directions and without primordial
	black holes, JCAP 0506:008,2005}[hep-ph/0310194].
\bibitem{hilltop-curvaton}
T. Matsuda,
{\it Hilltop curvatons, Phys.Lett.B659(2008)783} [arXiv:0712.2103].
\bibitem{dine-rand-thom}
M. Dine, L. Randall, S. D. Thomas,
{\it Supersymmetry breaking in the early universe,
Phys.Rev.Lett.75(1995)398} [hep-ph/9503303].
\bibitem{kin-fnl}
D. Babich, P. Creminelli and M. Zaldarriaga, 
{\it The Shape of non-Gaussianities, JCAP 0408:009,2004} 
[astro-ph/0405356].
\bibitem{cline-slow-extra}
N. Arkani-Hamed, S. Dimopoulos, N. Kaloper and J. March-Russell,
{\it Rapid asymmetric inflation and early cosmology in theories with
	submillimeter dimensions, Nucl.Phys.B567(2000)189}
	[hep-ph/9903224];
J. M. Cline,
{\it Inflation from extra dimensions,
Phys.Rev.D61(2000)023513} [hep-ph/9904495].
\bibitem{hilltop-inflation}
L. Boubekeur and D. H. Lyth,
{\it Hilltop inflation, JCAP 0507:010,2005}[hep-ph/0502047];
D. H. Lyth and T. Moroi,
{\it The masses of weakly-coupled scalar fields in the early universe,
JHEP 0405:004,2004} [hep-ph/0402174].
\bibitem{low-infla}
D. H. Lyth,
{\it Constraints on TeV scale hybrid inflation and comments on nonhybrid
	alternatives, Phys.Lett.B466(1999)85} [hep-ph/9908219];
A. M. Green and A. Mazumdar, 
{\it Dynamics of a large extra dimension inspired hybrid inflation model,
Phys.Rev.D65(2002)105022} [hep-ph/0201209];
R. N. Mohapatra, A. Perez-Lorenzana, and C. A. de S. Pires,
{\it Inflation in models with large extra dimensions driven by a bulk
scalar field, Phys.Rev.D62(2000)105030} [hep-ph/0003089];
T. Matsuda,
{\it Kaluza-Klein modes in hybrid inflation, Phys.Rev.D66(2002)107301}
[hep-ph/0209214];
T. Matsuda,
{\it Nontachyonic brane inflation, Phys.Rev.D67(2003)083519}
[hep-ph/0302035];
A. Mazumdar,
{\it New developments on embedding inflation in gauge theory and
	particle physics} [arXiv:0707.3350].
\bibitem{intersect-brane}
M. Cvetic, G. Shiu and  A. M. Uranga,
{\it Three family supersymmetric standard - like models from
	intersecting brane worlds, Phys.Rev.Lett.87(2001)201801}
[hep-th/0107143];
T. Matsuda,
{\it Comment on the stability of the Yukawa couplings and the
	cosmological problems of intersecting brane models,
Gen.Relativ.Gravit.37(2005)1297} [hep-ph/0309314].
\bibitem{fnl_from_PR_other}
K. Enqvist, A. Jokinen, A. Mazumdar, T. Multamaki and A. Vaihkonen,
{\it Non-Gaussianity from Preheating, Phys.Rev.Lett.94(2005)161301}
[astro-ph/0411394];
K. Enqvist, A. Jokinen, A. Mazumdar, T. Multamaki and A. Vaihkonen,
{\it Non-gaussianity from instant and tachyonic preheating,
JCAP {0503}, 010 (2005)} [hep-ph/0501076];
A. Jokinen and A. Mazumdar,
{\it Very Large Primordial Non-Gaussianity from multi-field: Application  
to Massless Preheating, JCAP {0604}, 003 (2006)} [astro-ph/0512368].
\bibitem{two-field}
  C. Gordon, D. Wands, B. A. Bassett and R. Maartens,
{\it Adiabatic and entropy perturbations from inflation,
Phys.Rev.D63(2001)023506} [astro-ph/0009131].
\end{thebibliography}
\end{document}